\documentclass[preprint,aps,nofootinbib]{revtex4}
\usepackage{graphicx}
\usepackage[dvips]{color}
\input epsf
\begin{document}

\newcommand{\be}{\begin{equation}}
\newcommand{\ee}{\end{equation}}
\def\bq{\begin{eqnarray}}
\def\eq{\end{eqnarray}}

\title{\bf Einstein's Theory of Gravity in the Presence of  Pressure: A Review}
\author{Ram Gopal Vishwakarma\footnote{Electronic Address: rvishwa@mate.reduaz.mx}}

\address{Unidad Acad$\acute{e}$mica de Matem$\acute{a}$ticas\\
 Universidad Aut$\acute{o}$noma de Zacatecas\\
 C.P. 98068, Zacatecas, ZAC,
 Mexico}

\begin{abstract}
The mysterious `dark energy' needed to explain the current observations, poses a serious confrontation between fundamental physics and cosmology.
The present crisis may be an outcome of the (so far untested) prediction of the general theory of relativity that the pressure of the matter source also gravitates. In this view, a theoretical analysis reveals some surprising inconsistencies and paradoxes faced by the energy-stress tensor (in the presence of pressure) which is used to model the matter content of the universe, including dark energy.


\end{abstract}
\maketitle

\section{\bf Introduction}

Recent observations indicate that the Universe is now in a phase of an accelerated expansion, the physical cause of which has become one of the most profound mysteries of science. 
In the framework of the relativistic theories of gravitation, this requires the inclusion of a term with the characteristics of repulsive gravity, dubbed as `dark energy' (specification of perfect fluid exerting a negative pressure), which accounts for about 75 percent of the total energy density in the universe. 
According to the `concordance model', the dark energy is the effect of the cosmological constant originally introduced by Einstein in some other context. However, the usual interpretation of this term as the zero-point energy of quantum fields leads to a dramatic disagreement between theory and observations by some 123 orders of magnitudes! It seems that even the most adventurous extensions of the standard model of particle physics are not capable to provide a satisfactory candidate for the dark energy particles. The present situation has compelled many theorists to look for some fundamental modifications in the laws of physics.

It should be noted that the peculiar property, negative pressure, of dark energy, is directly related to the prediction of the general relativity (GR) that pressure of the matter source also gravitates (in addition to its energy density). However, this prediction has not been tested in any experiment so far. 
Though the theory of GR has been tested to high precisions in several observations, all these tests have been limited to the pressureless cases only. The celebrated classical tests of GR consider empty space approximation ($\rho=p=0$). The same is true for the more precise test of GR through the observations of the radio pulsars which are rapidly rotating strongly magnetized neutron stars. This test assumes the neutron stars as point-like objects and look for the relativistic corrections to the Kepler's orbit (in the so called post-Keplerian parameters) by measuring the pulsar timing. The test does not even know the exact nature of the matter that pulsars and other neutron stars are made of.    

This is the time one should devise an experiment to test directly the prediction of GR that pressure of matter also gravitates. Here we examine this issue on the theoretical front and to our surprise, we find that the theory seems to suffer from some fundamental inconsistencies. Especially, the standard formulations of the energy-stress tensor seem to suffer from paradoxes and inconsistencies in the presence of pressure, as we shall see in the following.

\section{\bf Gravitational Effect of Pressure}

The theory of GR, the most remarkable combination of philosophical penetration, physical intuition and mathematical excellence, is regarded a highly successful theory of gravitation in terms of its agreement with experimental results and its ability to predict new phenomena.
One of the most novel aspects of this theory (of all the tensor field theories of gravitation, in fact) is its prediction that not only the energy density but the pressure of matter also causes gravitational attraction. 
In the Newtonian
framework, a positive pressure has a unique repulsive feature. 
Whereas 
in the relativistic equations of GR,
the pressure $p$ of the matter source enters algebraically with 
the same sign as the density $\rho$ of the matter source, implying that
 for physically realistic matter ($p>0$), pressure 
 adds to the energy density of the source field and hence contributes to the collapse of matter in the same way as does the energy density. This novel aspect of a pressure term in the Einsteinian gravitation is a purely relativistic effect resulting from the covariant character of the theory.

In order to provide an example of this {\it `gravitational effect of pressure'} in GR, one may consider the general relativistic model of a star where pressure plays a significant role in its hydrostatic equilibrium.
As the field surrounding an attracting point mass would be static and spherically symmetrical in nature, the
model of the star is given by the static spherically symmetric metric
\be
ds^2=e^{\nu(r)} c^2 dt^2 - e^{\lambda(r)} dr^2-r^2 d\theta^2-r^2 \sin^2\theta ~ d\phi^2\label{eq:metric}
\ee
applied to a system consisting of a perfect fluid
\be
T^{\alpha \beta}=(\rho c^2+p)\frac{dx^\alpha}{ds}\frac{dx^\beta}{ds}-p~g^{\alpha\beta},\label{eq:emtensor}
\ee
where $p$ is the proper hydrostatic (normal) pressure of the fluid and 
$\rho c^2$ is its proper energy density measured 
by an observer moving with the fluid ($dx^\alpha/ds=e^{-\nu/2}\delta^\alpha_0$). $c$ is the 
speed of light. After solving the Einstein field equations 
for (\ref{eq:metric}) and 
(\ref{eq:emtensor}) and eliminating $\lambda$ from the resulting equations, one
can write
\be
\frac{dp}{dr}+(\rho c^2+p)\frac{d\ln e^{\nu/2}}{dr}=0,\label{eq:rel}
\ee
which is the relativistic generalization of the Newtonian equation of hydrostatic equilibrium of the star: 
\be
\frac{dp}{dr}+\frac{GM(r)}{r^2}\rho=0.\label{eq:newton}
\ee
Note the additional contribution of $p$ to the term $\rho c^2$ in equation (\ref{eq:rel}).
The two terms on the l.h.s. of (\ref{eq:newton}) measure, respectively, two forces which are in equilibrium: an outward force (resulting from a positive pressure of a realistic matter with $\rho,~ p>0$) on a fluid element at a distance $r$ from the centre of the star, and an inward force on the fluid element exerted by the mass $M(r)=\int_0^r 4\pi r'^2 \rho(r')dr'$ of the star enclosed within the sphere of radius $r$.
 In the weak-field (Newtonian) 
limit, the time-time component of the metric tensor $g_{00}=e^\nu\rightarrow (1+2\psi/c^2)$ (where the Newtonian gravitational potential
$\psi=-GM/r$) and the second term of (\ref{eq:rel}) reduces to $GM\rho/r^2$ for $p<<\rho c^2$.

It may however be noted that equation (\ref{eq:rel}) contains, at the same time, two mutually contradictory aspects of $p$: the gradient of (a positive) $p$ provides an outward force resisting the collapse of matter (as in equation (\ref{eq:newton})), while the same $p$ added to $\rho c^2$ enhances the collapse.
Appearance of this additional attractive feature of a positive pressure term in GR  has remained a mystery of the theory which defies any proper explanation and one generally seeks refuge in the \textcolor{black}{{\it subtleties}} of GR while failing to provide any explanation.

 However, recalling that there always exists a local inertial reference frame in GR in which the {\it subtleties} of GR and gravity disappear locally, one may use it to understand this mysterious {\it `gravitational effect of pressure'} and find out its origin. This exercise however reveals that \textcolor{black}{the standard formulation of the energy-stress tensor of a perfect fluid is plagued with paradoxes}, as we shall see in the following.

\section{\bf A Paradox with the Energy-Stress Tensor}

In order to understand the gravitational effects of the pressure of a fluid, we examine the divergence of the energy-stress tensor (\ref{eq:emtensor}) which is famous for describing the mechanical behaviour of the fluid:
\be
T^{\alpha\beta}_{~ ~;\beta}=0,\label{eq:cons}
\ee
which follows from the Einstein equation through the Bianchi identities
($\{R^{\alpha\beta}-1/2 ~R g^{\alpha\beta}\}_{;\beta}=0$).
We choose a locally 
inertial frame of reference 
in which an element of the fluid is at rest in the neighbourhood of the observer,
at least momentarily (it is always possible to find such a coordinate system
in accordance with the principle of equivalence). 
In such a coordinate system, gravity disappears and GR reduces locally to the 
laws of SR (whence the expression of the energy-stress tensor is imported in GR through a general coordinate transformation). Hence we do not expect any {\it subtleties} of GR or gravity to 
creep in the analyses done by the observer. 

We note that in the chosen coordinate system, equation (\ref{eq:cons}) reduces 
to
\be
\frac{\partial T^{\alpha\beta}}{\partial x^\beta}=0.\label{eq:consf}
\ee
We consider $x^0\equiv ct, x^1\equiv x, x^2\equiv y, x^3\equiv z$.
We also recall that in this coordinate system, the first derivatives of the
$g_{\mu\nu}$ with respect to the coordinates vanish in the close neighbourhood
of the observer. However, the second derivatives will not vanish in general
(except for the special case of spacetime that is actually flat). Similarly,
the spatial components of the 4-velocity vector vanish in the  close 
neighbourhood of the observer, i.e.
\be
u_x\equiv\frac{dx}{d\tau}=0, ~u_y\equiv\frac{dy}{d\tau}=0, ~u_z\equiv\frac{dz}{d\tau}=0 ~ ~{\rm and}~~
\frac{dt}{d\tau}=1,\label{eq:vel}
\ee
where the proper time  $d\tau=ds/c$. However, the derivatives of the velocity will not be zero in general, except
for its temporal component, which will vanish in the chosen coordinates, as
we see in the following:
\[
ds^2=g_{\mu\nu}~dx^\mu dx^\nu \Rightarrow 
\]
\centerline{$
g_{00}\left(c\frac{dt}{ds}\right)^2+g_{11}\left(\frac{dx}{ds}\right)^2+....
+2g_{01}c\frac{dt}{ds}\frac{dx}{ds}+....
$}
\be
+2g_{12}\frac{dx}{ds}\frac{dy}{ds}+...=1,
\ee
which on differentiation gives
\be
\frac{\partial}{\partial x^\alpha}\left(\frac{dt}{d\tau}\right)=\frac{\partial}{\partial x^\alpha}\left(\frac{dt}{ds}\right)=0,\label{eq:velt}
\ee
by the virtue of the relations in  (\ref{eq:vel}) and by noticing that
$\partial g_{\mu\nu}/\partial x^\alpha=0$, as mentioned earlier.

We have now developed enough infrastructure to calculate (\ref{eq:consf})
for a fluid element in the neighbourhood of our observer. Following Tolman \cite{tolman}, we substitute
(\ref{eq:emtensor}) into (\ref{eq:consf}) for the case $\mu=1$ and use
(\ref{eq:vel}) and (\ref{eq:velt}) therein to obtain
\be
\frac{\partial p}{\partial x}+\left(\rho\textcolor{black}{+\frac{p}{c^2}}\right)\frac{du_x}{dt}=0,
\label{eq:mu1}
\ee
where $du_x/dt=du_x/d\tau=\partial u_x/\partial t$ (in the chosen coordinates)
is the acceleration of the fluid element in the x-direction. 
Equation (\ref{eq:mu1}), which is the relativistic analogue of the law of motion, is an important equation and provides
a clue of our problem. Equation  (\ref{eq:mu1}) actually signals towards an
inconsistency:
 the inertial mass of the fluid element has got an additional contribution! 
But, what is its source? Equation  (\ref{eq:mu1}), taken at the face value, reveals that $p$ should be carrying some kind of energy (density) as $p/c^2$ contributes to the inertial mass (density) of the fluid element. This also appears in tune with the generally made vague argument in an attempt to explain the peculiar behaviour of $p$ in GR that ``a positive pressure in GR  {\it somehow}  contains positive energy''.

However, \textcolor{black}{{\it it must be noted that the term $\rho c^2$ in} (\ref{eq:emtensor}), {\it as measured in the rest frame of the fluid, includes not only the rest mass of the individual particles of the fluid but also their kinetic energy, internal energy (for example, the energy of compression, energy of nuclear binding, etc.) and all other sources of mass-energy} \cite{MTW}.} \textcolor{black}{Thus, if $p$ `somehow' contains energy, it should be at the cost of violating the celebrated law of the conservation of energy}\footnote{Like many others, Tolman interprets the pressure in terms of the work done by the forces of stresses on the surroundings (see, for example, pages 66, 220 and 221 of \cite{tolman}). If this is correct (also see footnote 3), dividing $p$ by $c^2$ gives an equivalent mass density. However, the addition of this term to the mass density $\rho$ of the fluid element in equation (\ref{eq:mu1}), indicates a gain of energy by the element of the fluid (for realistic matter with $\rho>0$, $p>0$). This appears paradoxical with the conventional definition of the perfect fluid, 
which assumes that it is the collision of the constituent particles of the fluid, resulting from their
random motions, which generates pressure \cite{narlikar}.  
The work done by this pressure on the surroundings, which should be at the expense of the internal energy of the fluid element, would decrease (rather than increase) its net energy,
 according to the conventional wisdom about pressure.
 Consequently we should expect a decrease in the net inertia of the fluid element, contrary to what we find in (\ref{eq:mu1})! (One may try to resolve this paradox with a {\it hypothetical} matter with negative pressure. However, this would not resolve the paradox described in the following section, implying that the replacement of $p$ with $-p$ is not a solution.)  
} (now we cannot blame the mysteries of GR and gravity for an unexpected happening, as they are absent in the chosen coordinates). Though equation (\ref{eq:mu1}) is not an energy conservation equation, but that does not allow it to defy the law of conservation of energy.

{\it It should be noted that in the above-mentioned violation of the conservation of energy, there is no role of the notorious (pseudo) energy of the gravitational field which is absent here.}
 
It may be mentioned that the kinetic energy of the fluid particles is generally (mis)taken to be carried in $p$ and hence responsible for the additional contributions to their mass-energy  in equations (\ref{eq:rel}) and (\ref{eq:mu1}) 
 which is though not correct, as  we now know.
It may also be noted that the relativistic effect of increase of mass of the fluid particles with speed is taken into account in the term $\rho$, as is obvious (as we shall also see in section V), and cannot be considered responsible for the increase in the inertial mass of the fluid element in equation (\ref{eq:mu1}) through $p$. 

In the following, we describe another, already known, paradox emerging from the energy-stress tensor (\ref{eq:emtensor}).

\section{\bf Tolman Paradox}

 Tolman \cite{tolman} has derived, from the energy-stress tensor (\ref{eq:emtensor}), a  formula for the total energy of a fluid sphere in a quasi-static state: 
\be
U=\int (\rho c^2+3p)\sqrt{g_{00}}~dV,\label{eq:rmass} 
\ee
which holds for the quiescent states of temporary or permanent equilibrium. Here $dV$ is the proper spatial volume element of the fluid sphere. The quantity $U$ is a measure of the power of producing gravitational field by a fluid sphere in GR. Tolman himself has noticed a paradox (now known as Tolman Paradox \cite{ehlers}) related with the consequences of the  energy-stress tensor of the disordered radiation. By considering his formula (\ref{eq:rmass}) for the total energy of a fluid sphere, the paradox can be described as the following. The matter ($p=0$) at rest in a container exhibits a total mass $U/c^2$. However, converting the matter inside the container into disordered radiation ($p=\rho c^2/3$) would double the total mass, violating the conservation of $U$!

It would be honest to mention that Tolman's formula (\ref{eq:rmass}) makes use of the gravitational energy (through the use of $\sqrt{g_{00}}$), which is a controversial subject.
However, the paradox appears due to the factor $(\rho c^2+3p)$ in (\ref{eq:rmass}), and not due to the term $\sqrt{g_{00}}$.

\section{\bf The Origin of the Trouble}

In order to find out how pressure carries energy density in GR (i.e., the origin of the paradoxes) let us derive, from an action principle, the energy-stress tensor of the perfect fluid - the source term in Einstein's equation - which also seems to be the source of the trouble. Let us consider a particle ($i$-th) of rest mass $m_i$ in a small 4-volume of the fluid. Following Narlikar \cite{narlikar}, one can consider small variations of the type
\be
g_{\mu \nu}\rightarrow g_{\mu\nu}+\delta g_{\mu \nu}
\ee
in the action
\be
{\cal A}=\sum_i c m_i \int ds_i,
\ee
in order to write the energy-stress tensor as the following sum
\be
T^{\mu \nu}=\sum_i \frac{c^2}{E_i}P^\mu_{(i)}P^\nu_{(i)},\label{eq:action} 
\ee
over all those particles which cross the unit volume of the fluid. Here
\be
P^\mu_{(i)}= c m_i \left(\frac{dx^\mu}{ds}\right)_i 
\ee
is the 4-momentum of the $i$-th particle in a locally inertial coordinate system so that $E_i=c P^0_{(i)}$ is the energy of the particle, and  $ds_i^2=g_{\mu\nu}(dx^\mu)_i(dx^\nu)_i$ gives its proper time. By using the usual special-relativistic values for $P^\mu$ of a typical particle as the following


\[
P^0=\frac{mc}{\sqrt{1-v^2/c^2}}, ~~~P^1=\frac{mv_x}{\sqrt{1-v^2/c^2}}, 
\]
\be
P^2=\frac{mv_y}{\sqrt{1-v^2/c^2}},~~~P^3=\frac{mv_z}{\sqrt{1-v^2/c^2}},
\ee
one can identify the proper energy density $\rho c^2$ and the proper pressure $p$ of the fluid element (in the absence of any tangential shearing stresses) with the non-vanishing components of the tensor $T^{\mu \nu}$ measured by the local inertial observer in the following way:
\be
T^{00}=\sum_i\frac{m_i}{\sqrt{1-v_i^2/c^2}}c^2\approx\sum_i m_ic^2\left(1+\frac{v_i^2}{2c^2}\right)\equiv \rho c^2,\label{eq:rho}
\ee
\be
T^{11}=T^{22}=T^{33}=\frac{1}{3}\sum_i\frac{m_i}{\sqrt{1-v_i^2/c^2}}v_i^2\equiv p.\label{eq:p}
\ee
The factor $1/3$ comes from randomizing in all directions. For a general observer, to whom the fluid as a whole has a 4-velocity $dx^\alpha/ds$, the energy-stress tensor $T^{\mu \nu}$ takes the form given by (\ref{eq:emtensor}). \textcolor{black}{{\it Note that the relativistic effect of increase in mass of the particles with speed is already taken into account in $\rho$, as is clear from} (\ref{eq:rho}), {\it and cannot be considered responsible for the increase in the energy-mass density of the fluid element in equations} (\ref{eq:rel}) {\it and} (\ref{eq:mu1}).} In fact, this unexpected contribution to the energy-mass density is brought about by the term $p$ which is still non-zero ($\sum mv^2/3$) for non-relativistic motions of the particles. 

 Although in this basic derivation of $T^{\mu \nu}$ from the action principle, $p$ has the dimensions of the energy density in equation (\ref{eq:p}), we still cannot decipher how $p$ carries energy density in GR (without any source). In fact the belief, {\it that the pressure can be measured in terms of the kinetic energy density of the fluid}, stems from the Newtonian mechanics
 and is originally due to Daniel Bernoulli (1738). For example, the pressure of an ideal gas is derived in terms of its kinetic energy density. From the kinetic theory, the pressure arising from the force exerted by the gas molecules colliding with the walls of the container, can be derived as the following \cite{feynman}.

Consider a gas of N molecules, each of mass m, enclosed in a container. If a molecule makes an elastic collision  with the wall perpendicular to the $x$-direction with a speed $v_x$ and bounces off in the opposite direction with the same speed, the momentum lost by the molecule is $2mv_x$. As the time taken by the molecule to make the next collision with the wall is $2\ell/v_x$ (where $\ell$ is the length of the container), the force exerted on the wall in this collision is $mv_x^2/\ell$. The total force acting on the wall from all the molecules $=Nmv_{x_{\rm rms}}^2/\ell$, where $v_{x_{\rm rms}}=\sqrt{(v_{x_1}^2+v_{x_2}^2+....+v_{x_N}^2)/N}$ is the average (root-mean-square) speed of the collection of molecules. With an area $A$ of the wall, the pressure on the wall is then $=\rho v_{x_{\rm rms}}^2$, where $\rho\approx Nm/A\ell$ is the density of the gas (noting that the kinetic energy of the molecules, moving with non-relativistic speeds, would be insignificant compared with their rest mass energy). Assuming random speeds in all the directions, i.e., $v_{\rm rms}^2=v_{x_{\rm rms}}^2+v_{y_{\rm rms}}^2+v_{z_{\rm rms}}^2=3v_{x_{\rm rms}}^2$, the total average pressure yields 
\be
p=\frac{1}{3}\rho v_{\rm rms}^2,\label{eq:pNewton} 
\ee
which is the non-relativistic limit of equation (\ref{eq:p}). Generally, this result receives an interpretation that the pressure carries an energy density $=\frac{2}{3} \times$the kinetic energy density of the fluid.
\textcolor{black}{However, {\it if this is true, it leads to a Tolman's-like paradox in the non-relativistic case also: by converting some matter (put at rest inside a container) into gas}}\footnote{The energy required, if any, to convert the matter into gas would be accounted in the term $\rho c^2$.} \textcolor{black}{{\it or radiation, we would get extra energy from the pressure of the gas or radiation (as if in bonus!) violating the conservation of energy!}} 

What is the reason of this unexpected result appearing even in the Newtonian case (in the absence of any subtleties of GR or gravity)? A careful examination of the situation reveals that the \textcolor{black}{{\it pressure of a fluid is not its kinetic energy density}} (which is $\rho v_{\rm rms}^2/2$, not $\rho v_{\rm rms}^2/3$).
In fact, the interpretation of the pressure in terms of the (kinetic) energy density of the fluid supplies only half of the truth, i.e., it's magnitude only. It is clear from the derivation of equation (\ref{eq:pNewton}) that we just calculate the scalar magnitude of this vector quantity avoiding its direction. As the pressure in an ideal gas (perfect fluid) has a spherical symmetry, it is always possible to avoid its direction by choosing suitable (spherical polar) coordinates in which the pressure has a unique direction (in the direction of increasing $r$) everywhere. But that certainly does not mean that it is a scalar quantity.\footnote{Nor, the pressure is the work done
by the mechanical forces of stresses on the surroundings, as is clear from the derivation of equation (\ref{eq:pNewton}). The particles of the container make little or negligible movement, though the pressure of the gas inside the container may not be negligible. One can anyway estimate the magnitude of a force (pressure) in terms of the work that {\it would be done} by crossing unit length (volume), even though there is no work done in reality.} 
As soon as we realize that the pressure is a vector quantity, the paradox disappears from the Newtonian case. (Of course, we cannot imagine to include magnitudes of the momentum, or the angular momentum of the constituent particles of the fluid in its energy density!)  

In fact, \textcolor{black}{{\it interpreting the pressure as a scalar, is equivalent to assuming that in the presence of pressure, the fluid carries an additional amount of energy density}} (which though does not exist) since the dimensions of the pressure are that of the energy density [hence gradient of either of the two supplies force; that is why we do not face any trouble in interpreting the repulsive force which balances the gravitational pull in equation (\ref{eq:rel})]. \textcolor{black}{{\it It is this spurious energy density which is the cause of the paradoxes mentioned above in the} GR {\it and the Newtonian} {\it cases:}} the energy/mass density of the fluid in equations (\ref{eq:rel}), (\ref{eq:mu1}) and (\ref{eq:rmass}) seems to receive contribution from this illusionary energy density (that does not really exist!). 

 Though the realization of the pressure as a vector quantity removes the paradox from the Newtonian case, the problem is not so trivial in the GR where the pressure, like the energy density, has got to appear as components of a tensor as required by the covariant character of the theory. Though in GR, the pressure is realized primarily as components of a (second-rank) tensor in general, this feature is lost in the case of a fluid (in the absence of tangential shearing forces) and the tensor reduces to a diagonal form,  with the diagonal components identical (otherwise a rotation of the frame of reference would reveal presence of shear stress). This leaves the pressure $p$ in GR just a scalar and hence the paradoxes mentioned above remain there in GR.

\section{\bf Conclusion}

According to Einsteinian gravitation, it is not only the energy density that gravitates, but pressure also does so. This is so because
general relativity calls for the curvature of spacetime to be produced by
the mass-energy and pressure content of the matter in spacetime. This new aspect of pressure could not have been tested in any experiment so far as it is undetectably small in ordinary circumstances. However, the way this novel feature of gravitation imported by the pressure 
term appears in GR, leads to paradoxes and inconsistencies. In fact, a critical examination of the theory reveals 
 that in presence of the pressure, the source (for example, a perfect fluid) seems to carry additional energy density, though without any apparent source, hence defying the principle of conservation of energy and posing paradoxes. This unexpected fact, which is otherwise hidden in the subtleties of GR, is revealed by examining the conservation equation in an inertial system wherein GR and gravitation disappear locally.
It is then noted in equations (\ref{eq:rel}),  (\ref{eq:mu1}) and (\ref{eq:rmass}) that the mass/energy density receives contributions from an illusionary energy density (that does not really exist) attributed to the pressure term by violating the law of conservation of energy and hence giving rise to paradoxical situations.

It appears that the origin of the trouble lies in our erroneous belief that pressure is a scalar quantity, which becomes equivalent to assuming that in the presence of pressure, the fluid carries an additional amount of energy density (which though does not really exist). It is this spurious energy density which is the cause of the paradoxes, appearing not only in GR but in the Newtonian theory as well. The paradox disappears from the Newtonian theory as soon as we realize the pressure as a vector quantity, however the situation in GR is different where the pressure (like the energy density) has got to appear as components of a tensor, in order to maintain the covariant character of the theory. 

 It is generally argued that as the Euler- and mass balance- equations follow from the vanishing divergence of the energy-stress tensor, this should be regarded as a confirmation of the correctness of this tensor. It should however be noted that this happens only in the absence of the pressure term. In the presence of a high pressure or velocity, the relativistic laws differ from the classical ones \cite{adler}. 
There are a number of important situations in which pressure and energy density contribute comparably to the right-hand side of Einstein's equation, for example, the interior of a neutron star, the early radiation-dominated phase of the universe, the present accelerating phase of the universe. 
The predictions of Einstein's gravity, therefore, in these situations become suspect.  It is also obvious that all those theories of gravitation, which use the energy-stress tensor to represent matter (as is common in the relativistic theories of gravitation), are going to face this crisis. 

It may be mentioned that the precise observations of the CMB radiation by the COBE and then the WMAP satellites are generally interpreted as a complete confirmation of the hot big bang origin of the universe, and hence an early radiation dominated phase. 
However, the recent observations of the high energy cosmic rays with energies above a cut-off of $6\times10^{19}$ eV create doubts over the cosmological origin of the CMB \cite{CosmicRays}. It should be noted that the cosmic ray protons with energy above this cut-off cannot propagate through the CMB photons for a distance longer than 50 Mpc due to the energy loss process. As several events with energy above this cut-off have been observed, this points out towards a possibility of a local origin to the CMB. Additionally there are also alternative explanations of the CMB radiation in terms of the thermalized starlight from galaxies and clusters \cite{QSSC}.

Despite the remarkable success of GR, many researchers interpret the observations supporting the requirement of dark matter and dark energy as a failure of the theory. It seems that the validity of the theory is also questionable in the presence of a pressure term.
Finally, it may be mentioned that although the equations used in the above have been around for quite some time, the consequences (so far unknown) of the analysis are remarkable and far-reaching.

\bigskip

\noindent
``{\it In questions of science, the authority of a thousand is not worth
the humble reasoning of a single individual}''
\begin{flushright}
----- Galileo Galilei
\end{flushright}

\end{document}